\documentclass[12pt]{article}
\usepackage{graphicx}
\usepackage{amsmath}
\usepackage[utf8]{inputenc} 
\def\CPV{$CP$ violation\xspace}
\def\P{\ensuremath{P}\xspace}
\def\T{\ensuremath{T}\xspace}
\def\ct{\ensuremath{C_T}\xspace}
\def\ctb{\ensuremath{\bar{C}_T}\space}
\newcommand{\vecp}[1]{\ensuremath{\vec{\ptot}_{#1}}\xspace}
\def\At{\ensuremath{A_T}\xspace}
\def\Atb{\ensuremath{\bar{A}_T}\xspace}
\def\atodd{\ensuremath{a^{\T\text{-odd}}_{\CP}}\xspace}
\newcommand{\cpasym}[2]{\ensuremath{\frac{#1 - #2}{#1 + #2}}\xspace}

\def\onehalf		{\ensuremath{\frac{1}{2}}\xspace}
\def\xprime     {\ensuremath{x^{\prime}}\xspace}
\def\xprimesq   {\ensuremath{x^{\prime 2}}\xspace}
\def\yprime     {\ensuremath{y^{\prime}}\xspace}
\def\yprimesq   {\ensuremath{y^{\prime 2}}\xspace}

\def\agamma     {\ensuremath{A_{\Gamma}}\xspace}

\def\dzkk       {\decay{\Dz}{\Kp\Km}}
\def\dzpipi     {\decay{\Dz}{\pip\pim}}

\def\dkpicf     {\decay{\Dz}{\Km\pip}}
\def\dkpiws     {\decay{\Dz}{\Kp\pim}}

\def\dstarpmDpipm   {\decay{\Dstarpm}{\Dz\Ppi^{\pm}}}
\def\DztoDzb {\ensuremath{\Dz -{\kern -0.16em \Dzb}}\xspace}

\newcommand\pubnumber{WSU--HEP--XXYY}
\newcommand\pubdate{\today}

\def\bristol{H.H. Wills Physics Laboratory\\
University of Bristol, Bristol BS8 1TL, United Kingdom}
\def\support{\footnote{on behalf of the LHCb collaboration.}}
 
\def\Title#1{\begin{center} {\Large #1 } \end{center}}
\def\Author#1{\begin{center}{ \sc #1} \end{center}}
\def\Address#1{\begin{center}{ \it #1} \end{center}}

\newcommand\pubblock{\rightline{\begin{tabular}{l} \pubnumber\\
         \pubdate  \end{tabular}}}
\newenvironment{Abstract}{\begin{quotation}  }{\end{quotation}}
\newenvironment{Presented}{\begin{quotation} \begin{center} 
             PRESENTED AT\end{center}\bigskip 
      \begin{center}\begin{large}}{\end{large}\end{center} \end{quotation}}

\def\beq{\begin{equation}}
\def\eeq#1{\label{#1}\end{equation}}

\def\beqa{\begin{eqnarray}}
\def\eeqa#1{\label{#1}\end{eqnarray}}

\let\bar=\overbar

\def\D{{\cal D}}

\def\Dslash{\not{\hbox{\kern-4pt $D$}}}
\def\dslash{\not{\hbox{\kern-2pt $\del$}}}

\def\msb{{\bar{\ssstyle M \kern -1pt S}}}

\usepackage{ifthen} 
\newboolean{uprightparticles}
\setboolean{uprightparticles}{false} 
\usepackage{xspace} 
\usepackage{upgreek}

\def\lhcb {\mbox{LHCb}\xspace}

\ifthenelse{\boolean{uprightparticles}}%
{

 \def\Ppi         {\ensuremath{\uppi}\xspace}

 \def\PDelta      {\ensuremath{\Delta}\xspace}                 
 \def\PXi      {\ensuremath{\Xi}\xspace}                 
 \def\PLambda      {\ensuremath{\Lambda}\xspace}                 
 \def\PSigma      {\ensuremath{\Sigma}\xspace}                 
 \def\POmega      {\ensuremath{\Omega}\xspace}                 
 \def\PUpsilon      {\ensuremath{\Upsilon}\xspace}                 
 
 \def\PB      {\ensuremath{\mathrm{B}}\xspace}                 
                  
 \def\PD      {\ensuremath{\mathrm{D}}\xspace}

 \def\PK      {\ensuremath{\mathrm{K}}\xspace}

 \def\Pi      {\ensuremath{\mathrm{i}}\xspace}

}
{

 \def\Ppi         {\ensuremath{\pi}\xspace}

 \mathchardef\PDelta="7101
 \mathchardef\PXi="7104
 \mathchardef\PLambda="7103
 \mathchardef\PSigma="7106
 \mathchardef\POmega="710A
 \mathchardef\PUpsilon="7107
                  
 \def\PB      {\ensuremath{B}\xspace}                 
                  
 \def\PD      {\ensuremath{D}\xspace}

 \def\PK      {\ensuremath{K}\xspace}

 \def\Pi      {\ensuremath{i}\xspace}

}

\makeatletter
\ifcase \@ptsize \relax
  \newcommand{\miniscule}{\@setfontsize\miniscule{4}{5}}
\or
  \newcommand{\miniscule}{\@setfontsize\miniscule{5}{6}}
\or
  \newcommand{\miniscule}{\@setfontsize\miniscule{5}{6}}
\fi
\makeatother

\DeclareRobustCommand{\optbar}[1]{\shortstack{{\miniscule (\rule[.5ex]{1.25em}{.18mm})}
  \\ [-.7ex] $#1$}}

\def\pion   {{\ensuremath{\Ppi}}\xspace}

\def\pip    {{\ensuremath{\pion^+}}\xspace}
\def\pim    {{\ensuremath{\pion^-}}\xspace}

\def\kaon    {{\ensuremath{\PK}}\xspace}
  \def\Kbar    {{\kern 0.2em\overline{\kern -0.2em \PK}{}}\xspace}

\def\KorKbar    {\kern 0.18em\optbar{\kern -0.18em K}{}\xspace}

\def\Kp      {{\ensuremath{\kaon^+}}\xspace}
\def\Km      {{\ensuremath{\kaon^-}}\xspace}

  \def\Dbar    {{\kern 0.2em\overline{\kern -0.2em \PD}{}}\xspace}
\def\D       {{\ensuremath{\PD}}\xspace}

\def\DorDbar    {\kern 0.18em\optbar{\kern -0.18em D}{}\xspace}
\def\Dz      {{\ensuremath{\D^0}}\xspace}
\def\Dzb     {{\ensuremath{\Dbar{}^0}}\xspace}

\def\Dstarpm {{\ensuremath{\D^{*\pm}}}\xspace}

\def\B       {{\ensuremath{\PB}}\xspace}
\def\Bbar    {{\ensuremath{\kern 0.18em\overline{\kern -0.18em \PB}{}}}\xspace}

\def\BorBbar    {\kern 0.18em\optbar{\kern -0.18em B}{}\xspace}

\def\Bzb     {{\ensuremath{\Bbar{}^0}}\xspace}

\def\Bub     {{\ensuremath{\B^-}}\xspace}

\def\Bm      {{\ensuremath{\Bub}}\xspace}

  \def\Y#1S{\ensuremath{\PUpsilon{(#1S)}}\xspace}

\def\Lbar        {{\ensuremath{\kern 0.1em\overline{\kern -0.1em\PLambda}}}\xspace}
\def\LorLbar    {\kern 0.18em\optbar{\kern -0.18em \PLambda}{}\xspace}

\newcommand{\decay}[2]{\ensuremath{#1\!\to #2}\xspace}         

\def\to                 {\ensuremath{\rightarrow}\xspace}

\def\CP                {{\ensuremath{C\!P}}\xspace}

\def\AT#1     {\ensuremath{A_{\mathrm{T}}^{#1}}\xspace}           

\def\C#1      {\ensuremath{\mathcal{C}_{#1}}\xspace}                       
\def\Cp#1     {\ensuremath{\mathcal{C}_{#1}^{'}}\xspace}                    
\def\Ceff#1   {\ensuremath{\mathcal{C}_{#1}^{\mathrm{(eff)}}}\xspace}        
\def\Cpeff#1  {\ensuremath{\mathcal{C}_{#1}^{'\mathrm{(eff)}}}\xspace}       
\def\Ope#1    {\ensuremath{\mathcal{O}_{#1}}\xspace}                       
\def\Opep#1   {\ensuremath{\mathcal{O}_{#1}^{'}}\xspace}                    

\def\xprime     {\ensuremath{x^{\prime}}\xspace}
\def\yprime     {\ensuremath{y^{\prime}}\xspace}

\def\agamma     {\ensuremath{A_{\Gamma}}\xspace}
\def\dkpicf     {\decay{\Dz}{\Km\pip}}

\newcommand{\tev}{\ifthenelse{\boolean{inbibliography}}{\ensuremath{~T\kern -0.05em eV}\xspace}{\ensuremath{\mathrm{\,Te\kern -0.1em V}}}\xspace}
\newcommand{\gev}{\ensuremath{\mathrm{\,Ge\kern -0.1em V}}\xspace}
\newcommand{\mev}{\ensuremath{\mathrm{\,Me\kern -0.1em V}}\xspace}
\newcommand{\kev}{\ensuremath{\mathrm{\,ke\kern -0.1em V}}\xspace}
\newcommand{\ev}{\ensuremath{\mathrm{\,e\kern -0.1em V}}\xspace}
\newcommand{\gevc}{\ensuremath{{\mathrm{\,Ge\kern -0.1em V\!/}c}}\xspace}
\newcommand{\mevc}{\ensuremath{{\mathrm{\,Me\kern -0.1em V\!/}c}}\xspace}
\newcommand{\gevcc}{\ensuremath{{\mathrm{\,Ge\kern -0.1em V\!/}c^2}}\xspace}
\newcommand{\gevgevcccc}{\ensuremath{{\mathrm{\,Ge\kern -0.1em V^2\!/}c^4}}\xspace}
\newcommand{\mevcc}{\ensuremath{{\mathrm{\,Me\kern -0.1em V\!/}c^2}}\xspace}

\newcommand{\chisq}{\ensuremath{\chi^2}\xspace}

\def\gsim{{~\raise.15em\hbox{$>$}\kern-.85em
          \lower.35em\hbox{$\sim$}~}\xspace}
\def\lsim{{~\raise.15em\hbox{$<$}\kern-.85em
          \lower.35em\hbox{$\sim$}~}\xspace}

\def\ptot       {\mbox{$p$}\xspace}

\def\tell1  {TELL1\xspace}
\def\ukl1   {UKL1\xspace}

\makeatletter 
\renewcommand\@biblabel[1]{\enspace[#1]} 
\makeatother

\usepackage{cite}

\begin{document}

\begin{titlepage}
\pubblock

\vfill
\Title{Searches for \CPV in charm at LHCb}
\vfill
\Author{Paras P. Naik\support}
\Address{\bristol} 
\vfill
\begin{Abstract} 
LHCb has collected the world's largest sample of charmed hadrons.
This sample is used to search for direct and indirect \CPV
in charm. Recent and updated measurements from several decay modes 
are presented.  
\end{Abstract}
\vfill
\begin{Presented}
The 7th International Workshop on Charm Physics (CHARM 2015)\\
Detroit, Michigan, U.S.A., 18-22 May, 2015
\end{Presented}
\vfill
\end{titlepage}
\def\thefootnote{\fnsymbol{footnote}}
\setcounter{footnote}{0}
%

\section{Introduction}
Studies of \CPV are a primary goal of the LHCb Experiment
\cite{Alves:2008zz} at CERN. 
\CPV in charm decays is expected to be very small in the Standard Model
\cite{Bianco:2003vb,Grossman:2006jg}.
However, asymmetries at a few times $10^{-3}$ within the Standard Model cannot
be excluded \cite{Feldmann:2012js,Brod:2011re,Bhattacharya:2012ah}. 
A significant excess of
\CPV with respect to the theoretical predictions would be a signature of physics
beyond the Standard Model. The study of \CPV in singly Cabibbo-suppressed (SCS)
charm decays is uniquely sensitive to physics beyond the Standard Model, in
particular through new contributions to $\Delta C = 1$ strong penguin and chromomagnetic dipole operators~\cite{Grossman:2006jg}. 
The analysis of the
following decay channels allows us several opportunities to probe both
integrated and localized \CPV in different regions of decay phase space.
Large \CPV has been observed in regions of
meson decay phase space not associated to a resonance, further
motivating localized searches~\cite{Aaij:2014iva}.
To date, no \CPV has been observed in the charm quark
system.

\section{Searches for direct \CPV in charm}

\subsection{${D^+_{(s)}} \to K^0_S h^+$}
\label{twoone}

The measured
time-integrated \CP asymmetry for a meson ($Q$) decay to a final state $f$ can
be given by:
\begin{equation}
A_{\rm{raw}}(Q \to f)=\frac{N(Q \to f)-N(\overline{Q} \to
\overline{f})}{N(Q \to f)+N(\overline{Q} \to\overline{f})},
\end{equation}
where $N$ indicates the number of reconstructed events of a given decay after
background subtraction.
We study the decays $D^+_{(s)}\to K^0_S h^+$, where $h = \pi~{\rm{or}}~K$. \CPV
is only expected to be measurable in the SCS subset of these decays.  The analysis
is performed using promptly produced $D$ mesons from the full
$3~{\rm{fb}}^{-1}$ LHCb Run~1 data sample, which consists of $1~{\rm{fb}}^{-1}$
of proton-proton collision data taken in 2011 at $\sqrt{s} =$~7~TeV, and
$2~{\rm{fb}}^{-1}$ of proton-proton collision data taken in 2012 at $\sqrt{s} =$~8~TeV.

These measured (or ``raw'') asymmetries are a sum of several independent
asymmetries: 
\begin{equation}
A_{\rm{raw}}(K^0_S h^+) = A_{CP}(K^0_S h^+)+A_{\rm{P}}(D^+_{(s)})+A_{\rm{D}}(h^+) +
A_{CP/{\rm{int}}},
\end{equation}
where $A_{CP}$ is the physical \CP asymmetry, $A_{\rm{P}}$ is
the production asymmetry of the decay particle,  $A_{\rm{D}}$ is the detection
asymmetry of the charged daughter hadron, and 
$A_{CP/{\rm{int}}}$
represents the 
 detection asymmetry between a $K^0$
and a ${\bar{K}{}^0}$, due to 
the 
presence of mixing and \CPV in the kaon
system
and the different interaction rates of 
$K^0$ and $\bar{K}{}^0$ in the detector material~\cite{Aaij:2014gsa}. 

The quantities of interest can then be determined by forming differences that
result in the cancellation of production and detection asymmetries: 
\begin{equation}
A_{CP}(D^+_S\to K^0_S\pi^+) = A_{\rm{raw}}(D^+_s\to K^0_S\pi^+)-A_{\rm{raw}}(D^+_s\to \phi \pi^+)
\label{three}
\end{equation}
and
\begin{eqnarray}
A_{CP}(D^+\to K^0_S K^+) &=& (A_{\rm{raw}}(D^+\to K^0_S K^+)-A_{\rm{raw}}(D^+_s \to K^0_S K^+)) \\ \nonumber
&- &(A_{\rm{raw}}(D^+\to K^0_S\pi^+) - A_{\rm{raw}}(D^+_s\to \phi \pi^+)),
\label{four}
\end{eqnarray}
where we assume the Standard Model expectation of no
$CP$ violation in the Cabibbo-favored (CF) $D_s^+ \to \phi \pi^+$ decay mode.

As a cross check, a double difference of asymmetries largely insensitive to
production and instrumental asymmetries may also be formed:
\begin{eqnarray}
\label{five}
A_{CP}^{DD} &=& ((A_{\rm{raw}}(D^+_s\to K^0_S\pi^+) - (A_{\rm{raw}}(D^+_s\to K^0_S K^+)) \\ \nonumber
&-& ((A_{\rm{raw}}(D^+\to K^0_S\pi^+) - (A_{\rm{raw}}(D^+\to K^0_S K^+))\\ \nonumber
&\approx& A_{CP}(D^+_s\to K^0_S\pi^+) + A_{CP}(D^+\to K^0_S K^+).
\end{eqnarray}
In Equations~\ref{three}-\ref{five} above, the neutral kaon asymmetry is not
shown, but is accounted for in the final results.
As the production and detection asymmetries depend on 
the kinematic distributions, weights are assigned to the 
candidates such that the kinematic distributions are equalized. 

The analysis is performed separately for 2011 and 2012 data, and for both
LHCb magnet polarities; consistent results are obtained. 
The combined results~\cite{Aaij:2014qec}  
\begin{equation}
A_{CP}(D^+\to K^0_S K^+) = (0.03\pm0.17{(\rm{stat})}\pm0.14{(\rm{syst})})\%,
\end{equation}
\begin{equation}
A_{CP}(D^+_s\to K^0_S\pi^+) = (0.38\pm0.46{(\rm{stat})}\pm0.17{(\rm{syst})})\%,
\end{equation}
and
\begin{equation}
A_{CP}(D^+\to K^0_S K^+) + A_{CP}(D^+_s\to K^0_S\pi^+) =
(0.41\pm0.49{(\rm{stat})}\pm0.26{(\rm{syst})})\%,
\end{equation}
show no indication of \CPV. These 
are the most precise
measurements of these quantities.

\subsection{$\Dz \to h^+ h^-$ ($\Delta A_{CP}$)}  

A search for a time-integrated \CP asymmetry in 
$\Dz\to h^+h^-$ decays is
performed using the full LHCb Run~1 data sample ($3~{\rm{fb}}^{-1}$). 
The flavor of the
initial $\Dz$ state is tagged by the charge 
of the muon in semileptonic $B\to\Dz\mu^-\overline{\nu}_{\mu}X$ decays, where
$B$ is a $\Bm$ or $\Bzb$.
The measured asymmetry for tagged neutral $D$ mesons to a $CP$ conjugate final
state $f$ is given by:
\begin{equation}
A_{\rm{raw}}(f)=\frac{N(B\to\Dz\mu^-X')-N(B\to\Dzb\mu^+X')}{N(B\to\Dz\mu^-X')+N(B\to\Dzb\mu^+X')},
\end{equation}
where  
$N$ indicates
the number of reconstructed events of a given decay after background
subtraction
 and $X'$ represents the set of undetected final state
particles from the semileptonic $B$ decay.

The measured asymmetry is a sum of the physical \CP asymmetry ($A_{CP}(f)$),
the production asymmetry ($A_{\rm{P}}(B)$) and detection asymmetry
($A_{\rm{D}}(\mu)$):
\begin{equation}
A_{\rm{raw}}(f)=A_{CP}(f)+A_{\rm{P}}(B)+A_{\rm{D}}(\mu).
\end{equation}
An experimentally robust variable may be constructed, $\Delta
A_{CP}$, by taking the difference of the measured 
asymmetries in $\Dz\to K^+K^-$ and $\Dz\to \pi^+\pi^-$ decays:
\begin{equation}
\Delta A_{CP} = A_{\rm{raw}}(KK)-A_{\rm{raw}}(\pi\pi),
\end{equation}
as the production and the muon detection asymmetries 
cancel to first
order. 

Alternatively, for extracting $A_{CP}(KK)$, the detection 
and production 
asymmetries can be measured using CF 
$B\to\Dz(\to K^-\pi^+)\mu^-X$ decays where no \CPV is expected. 
An additional
detection asymmetry, $A_{\rm{D}}(K\pi)$, arises due to the 
different 
interaction rates of the charged kaon with matter. 
To remove this asymmetry,
the CF control channels  $D^+\to
K^-\pi^+\pi^+$ and $D^+\to \bar{K}{}^0\pi^+$ (for which we assume negligible
$CP$ violation) are used. 
In the $D^+\to \bar{K}{}^0\pi^+$ decays, 
$A_{CP/{\rm{int}}}$ (see Section~\ref{twoone}) is estimated from simulation
and subtracted from the measured asymmetry. 
Once $\Delta A_{CP}$ and 
$A_{CP}(KK)$ are measured, the individual asymmetry 
$A_{CP}(\pi \pi) = \Delta A_{CP} - A_{CP}(KK)$ can be computed.

The analysis is performed separately for the 2011 and the 2012 data, 
and for the two magnet polarities; consistent results are 
obtained. The combined results for the asymmetries~\cite{Aaij:2014gsa}, 
\begin{equation}
\Delta A_{CP} =(+0.14\pm0.16{(\rm{stat})}\pm0.08{(\rm{syst})})\%,
\end{equation}
\begin{equation}
A_{CP}(KK)=(-0.06\pm0.15{(\rm{stat})}\pm0.10{(\rm{syst})})\%,
\end{equation}
and, with correlation $\rho=0.28$,
\begin{equation}
A_{CP}(\pi\pi)=(-0.20\pm0.19{(\rm{stat})}\pm0.10{(\rm{syst})})\%,
\end{equation}
are compatible with \CP conservation. 

The $A_{CP}(hh)$ asymmetries 
are the most precise measurement of these individual asymmetries to 
date. The precision of $\Delta A_{CP}$ is comparable to the 
preliminary result for $\Delta A_{CP}$ measured using prompt 
$\Dz$ decays reconstructed in 1~fb$^{-1}$ of LHCb Run~1
data from 2011~\cite{LHCb:2013dka}.

\subsection{$\Dz \to \pi^+ \pi^- \pi^0$ (Energy Test)}

The energy test~\cite{Williams:2011cd} is an unbinned 
model-independent
statistical method than can be used to search for time-integrated \CPV 
in $\Dz \to \pi^-\pi^+\pi^0$ decays. This 
method relies on the comparison of 
$\Dz$ and $\Dzb$ flavor samples and 
is sensitive to \CPV localized in the phase-space of the 
multi-body final state.

The flavor of the prompt $\Dz$ is tagged by the charge 
of the slow pion in the decay $D^* \to \Dz \pi_s$. For 
the reconstruction of the $\Dz$, 
two categories of reconstructed neutral pions are used:
pions
for
which
both
final
state
photons
are
reconstructed
separately
(resolved
pions),
as
well
as
pions
that
have
higher
momentum
(typically
$p_T > 2~{\rm{GeV}} / c$)
and
thus
a
smaller
opening
angle
of
the
two
photons
(merged
pions).

The
energy test is used to assign a $p$-value 
for a non-zero \CPV hypothesis. 
In this method, a test statistic $T$ is 
used to compare the average distances based on the metric 
function $\psi$.
The test statistic is defined as
\begin{equation}
T = \sum_{i,j>i}^{n}\frac{\psi_{ij}}{n(n-1)}
 + \sum_{i,j>i}^{\overline{n}}\frac{\psi_{ij}}{\overline{n}(\overline{n}-1)}
 - \sum_{i,j}^{n,\overline{n}}\frac{\psi_{ij}}{n\overline{n}} ,
\label{eqn:T}
\end{equation}
and the metric function $\psi_{ij}\equiv\psi(d_{ij})=e^{-d_{ij}^2/2\sigma^2}$ is chosen as a Gaussian function with a tunable parameter 
$\sigma$.
$T$ compares the average distances of pairs of events belonging to two samples of opposite flavor. The normalization factor removes the impact of global asymmetries. 
The distance between two points in phase space is given by 
$d_{ij}=(m_{12}^{2,j}-m_{12}^{2,i},m_{23}^{2,j}-m_{23}^{2,i},m_{13}^{2,j}-m_{13}^{2,i})$, 
where the subscripts \{$1,2,3$\} indicate the final-state particles.
 
If there is \CPV, $T$ is expected to be larger than zero.
This 
technique calculates a $p$-value 
under the hypothesis of $CP$ symmetry by comparing the nominal 
$T$ value observed in data to a distribution of $T$ values 
obtained from permutation samples, where the flavor of the 
$\Dz$ is 
randomly reassigned to simulate samples without \CPV. 
The $p$-value for the no \CPV hypothesis is obtained as the 
fraction of permutation $T$ values greater than the nominal 
$T$ value in the data.

We study the data sample of $2~{\rm{fb}}^{-1}$ collected by
LHCb during 2012, and find
a $p$-value of $(2.6\pm0.5)\times10^{-2}$~\cite{Aaij:2014afa}. 
This result is based on
1000 permutations.
The data sample has been split according to various criteria to test the stability of the results.
Analyses of sub-samples with opposite magnet polarity, with different trigger configurations, and with fiducial 
selection requirements removing areas of high local asymmetry of the tagging soft 
pion from the $D^{*+}$ decay 
all provide consistent results. Various checks have been performed to ensure
there are no asymmetries arising from background events or detector related asymmetries.
Varying the metric parameter $\sigma$
results in similar $p$-values at the $10^{-2}$ level. 

This analysis provides the world's best sensitivity from a single experiment to
local \CPV in $\Dz \to \pi^-\pi^+\pi^0$ decays.

\subsection{$\Dz \to K^+ K^- \pi^+ \pi^-$ (Triple product asymmetries)} 

Multi-body \PD decays are sensitive to 
\CPV due to the interference of several 
resonances across the multi-body phase space. 
In $\Dz \to K^+ K^- \pi^+ \pi^-$ decays, 
triple products of final state particle momenta in the 
\Dz rest frame, defined as 
\mbox{$\ct \equiv \vecp{\Kp} \cdot \left(\vecp{\pip} \times \vecp{\pim}\right)$} 
and 
\mbox{$\ctb \equiv \vecp{\Km} \cdot \left(\vecp{\pim} \times \vecp{\pip}\right)$}
for \Dz and \Dzb mesons, respectively,
are quantities which are \P-odd (and \T-odd).
The decay rate asymmetries:
\begin{equation}
\At  \equiv \cpasym{\Gamma(\ct > 0)}{\Gamma(\ct < 0)},~~
\Atb  \equiv \cpasym{\Gamma(-\ctb > 0)}{\Gamma(-\ctb < 0)}
\end{equation}
are thus sensitive to \CPV. However, final state 
interactions significantly alter the measured asymmetries, and so 
the difference $\atodd \equiv \onehalf (\At - \Atb)$ is used 
to access the \CP asymmetry of the \Dz meson by canceling 
final state interaction effects.
  
The observable \atodd is, by definition, insensitive to 
production and detection 
asymmetries, and is thus very robust against systematic uncertainties.
Using the full LHCb Run~1 data sample (3~fb$^{-1}$), the phase space
integrated measurements of the asymmetries are~\cite{Aaij:2014qwa}:
\begin{eqnarray}
  \At &= (\ensuremath{-}7.18 \pm 0.41 (\text{stat}) \pm 0.13 (\text{syst}))
  \%,\\ 
  \Atb &= (\ensuremath{-}7.55 \pm 0.41 (\text{stat}) \pm 0.12 (\text{syst}))
  \%, 
  \end{eqnarray}
  and
  \begin{eqnarray}
  \atodd &= (\ensuremath{~~\,}0.18 \pm 0.29 (\text{stat}) \pm 0.04  (\text{syst})) \%.
\end{eqnarray}
These results show no evidence of global \CPV in these decays, but 
achieve a significant improvement in precision 
over previous measurements~\cite{Amhis:2014hma}.

The same measurements are also performed in 
32 bins of  
the five-dimensional
phase
space.
The
asymmetries are extracted in each bin  
and 
\atodd calculated. A \chisq test for consistency 
across the phase space is performed, yielding a $p$-value 
of \mbox{74\%}. Thus there is no evidence for direct local \CPV in this decay.

Similarly, binning 
in the decay time of the \Dz candidates and 
performing the same test gives sensitivity to 
indirect \CPV. This yields a $p$-value of 
\mbox{72\%}; no evidence for indirect \CPV was found.

\section{Searches for indirect \CPV in charm}

\subsection{$\Dz \to h^+ h^-$ ($A_\Gamma$)}

\begin{figure}[t]
\centering
\includegraphics[width=0.47\textwidth]{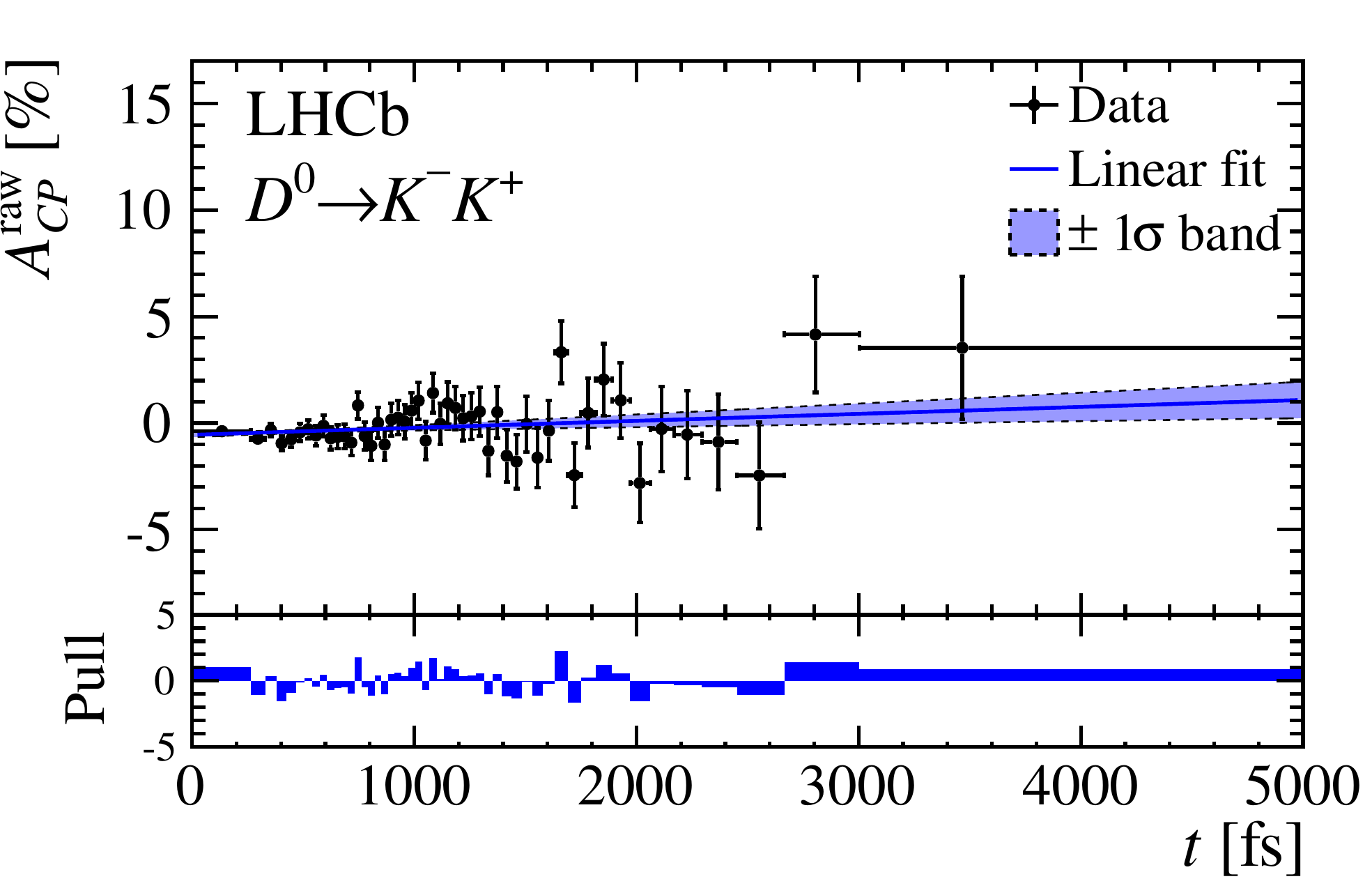}
\includegraphics[width=0.47\textwidth]{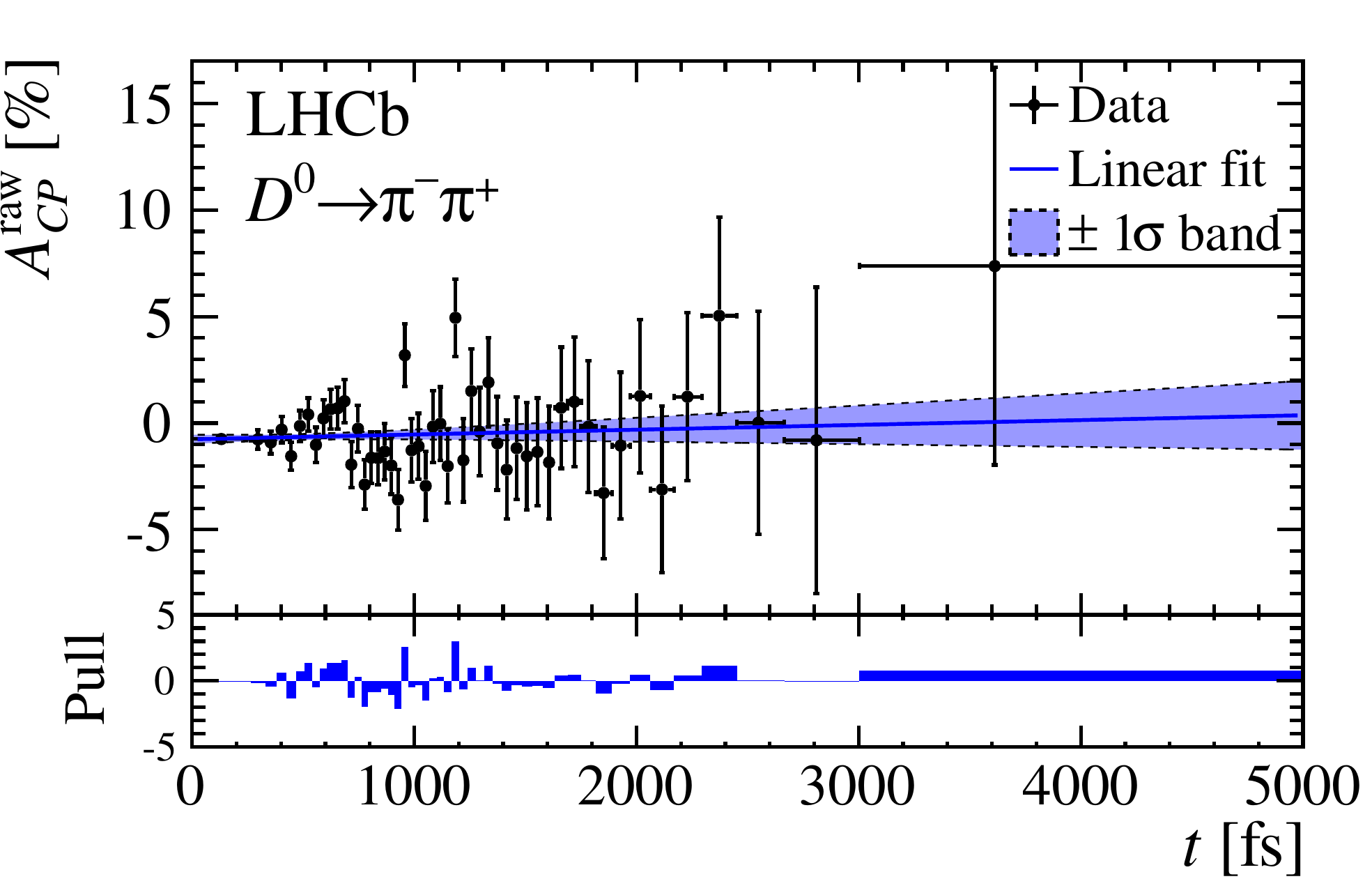}
\caption{Raw \CP asymmetry as a function of decay time for \dzkk decays (left) 
  and of \dzpipi decays (right), and corresponding pull
  plots~\cite{Aaij:2015yda}.}
\label{AGammaSL}
\end{figure}

A measurement of the indirect \CPV in \Dz~mixing can be
performed in the study of two-body hadronic charm decays to \CP
eigenstates (\dzkk or \dzpipi).   It can be evaluated by the 
asymmetry in 
the effective lifetimes ($\tau$) of flavor-tagged decays 
and expressed by the following equation with the assumption 
of negligible
direct \CP-violating contribution:
\begin{equation}
\agamma=\frac{\tau  (\Dzb \rightarrow f)-\tau (\Dz \rightarrow f)}
                 {\tau(\Dzb \rightarrow f)+\tau (\Dz \rightarrow f)}
\approx \frac{1}{2}\: A_m\: y\:  \cos \phi - x\: \sin \phi 
\end{equation}
where $A_m$ is defined by  $|q/p|^{\pm 2} \approx 1 \pm A_m$. 
 A measurement of 
\agamma~differing significantly
from zero is a  manifestation of indirect \CPV as it requires
a non-zero value for $A_m$ or $\phi = {\rm{arg}}(q/p)$. 

A measurement of \agamma~at \lhcb is performed using 3~fb$^{-1}$ of data from
the full LHCb Run~1 data sample~\cite{Aaij:2015yda}.  
In this analysis, $D^0$ decays are obtained from 
$B\to\Dz\mu^-\overline{\nu}_{\mu}X$ decays and
the neutral $D$ meson is tagged by the charge of the muon.
$A_\Gamma$ is determined through a \chisq fit to the time-dependent raw \CP
asymmetry $A^{\rm{\rm{raw}}}_{\CP}(t)$:
\begin{equation}
A^{\rm{\rm{raw}}}_{\CP}(t) \approx A_{\rm{D}} - A_\Gamma\frac{t}{\tau}.
\end{equation}
The raw \CP asymmetry is determined from simultaneous fits to 
$m_{h^+h^-}$ 
in
bins of decay time for both \Dz and \Dzb samples, as seen in
Figure~\ref{AGammaSL}.

Systematic
uncertainties include uncertainty in the $D^0$ mass fit model, the decay time
resolution, mixing in the beauty quark system, detection and production
asymmetries, and the time-dependent efficiency. The main contributions to the
systematic uncertainty result from the mistag probability (the probability that
the neutral $D$ meson was tagged with the wrong flavor) and the mistag
asymmetry (differences in positive and negative muon mistag probability), which
are studied using a control sample of $D^0 \to K^-\pi^+$ decays. The mistag
asymmetry is the dominant contribution to the systematic uncertainty.

We find:
\begin{eqnarray}
A_\Gamma(KK) &=& (-0.134 \pm 0.077~^{+0.026}_{-0.034})\% \\
A_\Gamma(\pi\pi) &=& (-0.092 \pm 0.145~^{+0.025}_{-0.033})\%,
\end{eqnarray}
where the uncertainties are statistical and systematic, respectively. This
measurement is consistent with no indirect \CPV in the charm system, and
has comparable precision to LHCb's previous result using neutral $D$ mesons from
prompt $D^*$ decays obtained from 1~fb$^{-1}$ of LHCb Run~1 data from 2011~\cite{Aaij:2013ria}.

\subsection{“Wrong-sign” $\Dz \to K \pi$ (\CPV via mixing)}

At LHCb, charm mixing parameters are
determined by the time-dependent 
ratio of \dkpiws (``wrong''
sign, WS) to \dkpicf (``right sign'', RS) decay rates. 
The RS decay
rate is dominated by its CF amplitude. 
The WS rate arises
from the interfering amplitudes of the doubly 
Cabibbo-suppressed decay (DCS) 
and the CF decay following \DztoDzb oscillation.

Assuming no \CPV and small charm mixing parameters 
($|x|$ and $|y|$ $\ll
1$), this ratio is:
\begin{equation}
 R(t) \approx R_D + \sqrt{R_D} \yprime \frac{t}{\tau}
 +\frac{\xprimesq+\yprimesq}{4} \left( \frac{t}{\tau} \right)^2
\end{equation}
where $\xprime=x \cos \delta + y \sin \delta$, $\yprime=y \cos \delta - x
\sin \delta$, $R_D$ is the ratio of suppressed-to-favored decay
rates, and $\delta$ is the strong phase difference between the DCS decays 
and the CF decays
$\left( \mathcal{A}\left(\dkpiws \right)/\mathcal{A}\left(\dkpicf \right)=- \sqrt{R_D} e^{-i
  \delta}\right)$.

We can search for \CPV in these decays by comparing
the time-dependent ratios evaluated separately for 
\Dz~and \Dzb. A difference in the $R_D$ parameter for \Dz and \Dzb
would be a sign of direct \CPV, while a difference in $\xprimesq$
and $\yprime$ parameters would imply indirect \CPV
($|q/p|\neq 1$ or $\phi\neq 0$). 
The data are fit under three hypotheses:
\CP is conserved, only an indirect \CP contribution is allowed, and  
both direct and indirect \CP contributions are allowed.

\begin{figure}[t]
\centering
\includegraphics[width=\textwidth]{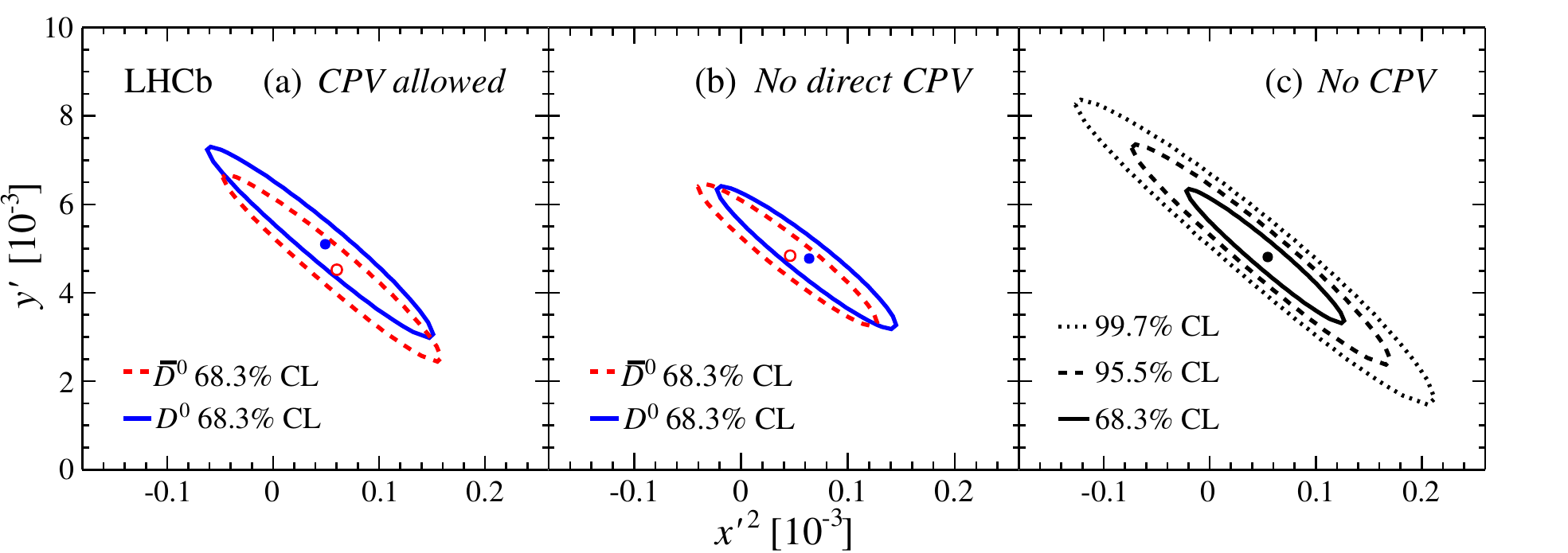}
\caption{Two-dimensional confidence regions in the ($\xprimesq$, $\yprime$)
  plane obtained (a) allowing both direct and indirect \CPV, (b) assuming no
  direct \CPV and (c) assuming \CP conservation. The solid (dashed)
  curves in (a) and (b) indicate the contours of the mixing parameters
  associated with \Dz~(\Dzb) decays.  The solid, dashed and dotted
  curves in (c) indicate the contours of \CP-averaged mixing parameters
  at $68.3\%$, 
  $95.5\%$ 
  and $99.7\%$ 
  C.L.
  The best-fit value is shown
  with a point~\cite{Aaij:2013wda}.}
\label{fig3}
\end{figure}

The full LHCb Run~1 data sample (3~fb$^{-1}$) is used to perform these
measurements.  
The sample is obtained from \dstarpmDpipm decays, 
which allow the determination of the flavor of the
neutral $D$ meson.
Assuming \CP conservation, the mixing parameters are measured to be $R_D=
(3.568\pm 0.058\pm 0.033)\cdot 10^{-3}$, $\yprime = (4.8 \pm 0.8\pm
0.5)\cdot 10^{-3}$ and $\xprimesq =( 5.5 \pm 4.2 \pm 2.6)\cdot 10^{-5}$
where the first uncertainty is statistical and the second
systematic~\cite{Aaij:2013wda}.
In the other scenarios, the direct \CP asymmetry is $A_D = (-0.7 \pm 1.9)\%$, 
and the magnitude of $q/p$ is determined to be $0.75 < | q/p | < 1.24$ and
$0.67 <| q/p |<1.52$ at $68.3\%$ and $95.5\%$ confidence level (C.L.), 
respectively.  The results of the 
measurements of mixing parameters, and of the confidence regions, are shown in
Figure~\ref{fig3} for the three scenarios. The results are compatible with \CP conservation and
provide the most stringent bounds on the parameter $| q/p |$ from a
single experiment.

\begingroup    
\raggedright           
\bibliographystyle{unsrt}    
\bibliography{charm2015_ParasNaik_bib}   

\begin{thebibliography}{10}

\bibitem{Alves:2008zz}
A.~Alves et~al.
\newblock {The LHCb Detector at the LHC}.
\newblock {\em JINST}, 3:S08005, 2008.

\bibitem{Bianco:2003vb}
S.~Bianco, F.L. Fabbri, D.~Benson, and I.~Bigi.
\newblock {A Cicerone for the physics of charm}.
\newblock {\em Riv.Nuovo Cim.}, 26N7:1--200, 2003.

\bibitem{Grossman:2006jg}
Y.~Grossman, A.L. Kagan, and Y.~Nir.
\newblock {New physics and $CP$ violation in singly Cabibbo suppressed $D$
  decays}.
\newblock {\em Phys.Rev.}, D75:036008, 2007.

\bibitem{Feldmann:2012js}
T.~Feldmann, S.~Nandi, and A.~Soni.
\newblock {Repercussions of Flavour Symmetry Breaking on $CP$ Violation in
  $D$-Meson Decays}.
\newblock {\em JHEP}, 1206:007, 2012.

\bibitem{Brod:2011re}
J.~Brod, A.L. Kagan, and J.~Zupan.
\newblock {Size of direct $CP$ violation in singly Cabibbo-suppressed $D$
  decays}.
\newblock {\em Phys.Rev.}, D86:014023, 2012.

\bibitem{Bhattacharya:2012ah}
B.~Bhattacharya, M.~Gronau, and J.L. Rosner.
\newblock {$CP$ asymmetries in singly-Cabibbo-suppressed $D$ decays to two
  pseudoscalar mesons}.
\newblock {\em Phys.Rev.}, D85:054014, 2012.

\bibitem{Aaij:2014iva}
R.~Aaij et~al.
\newblock {Measurements of $CP$ violation in the three-body phase space of
  charmless $B^{\pm}$ decays}.
\newblock {\em Phys.Rev.}, D90(11):112004, 2014.

\bibitem{Aaij:2014gsa}
R.~Aaij et~al.
\newblock {Measurement of $CP$ asymmetry in $D^0 \rightarrow K^- K^+$ and $D^0
  \rightarrow \pi^- \pi^+$ decays}.
\newblock {\em JHEP}, 07:041, 2014.

\bibitem{Aaij:2014qec}
R.~Aaij et~al.
\newblock {Search for $CP$ violation in $D^{\pm}\rightarrow K^0_{\mathrm{S}}
  K^{\pm}$ and $D^{\pm}_{s}\rightarrow K^0_{\mathrm{S}} \pi^{\pm}$ decays}.
\newblock {\em JHEP}, 10:25, 2014.

\bibitem{LHCb:2013dka}
R.~Aaij et~al.
\newblock {A search for time-integrated $CP$ violation in $D^0\rightarrow
  K^-K^+$ and $D^0\rightarrow \pi^-\pi^+$ decays}.
\newblock {\em LHCb-CONF-2013-003}, 2013.

\bibitem{Williams:2011cd}
M.~Williams.
\newblock {Observing $CP$ Violation in Many-Body Decays}.
\newblock {\em Phys. Rev.}, D84:054015, 2011.

\bibitem{Aaij:2014afa}
R.~Aaij et~al.
\newblock {Search for $CP$ violation in $D^0\to\pi^-\pi^+\pi^0$ decays with the
  energy test}.
\newblock {\em Phys. Lett.}, B740:158, 2015.

\bibitem{Aaij:2014qwa}
R.~Aaij et~al.
\newblock {Search for $CP$ violation using $T$-odd correlations in $D^0 \to
  K^+K^-\pi^+\pi^-$ decays}.
\newblock {\em JHEP}, 10:005, 2014.

\bibitem{Amhis:2014hma}
Y.~Amhis et~al.
\newblock {Averages of $b$-hadron, $c$-hadron, and $\tau$-lepton properties as
  of summer 2014}.
\newblock {\em arXiv:1412.7515}, 2014.

\bibitem{Aaij:2015yda}
R.~Aaij et~al.
\newblock {Measurement of indirect $CP$ asymmetries in $D^0\rightarrow K^-K^+$
  and $D^0\rightarrow \pi^-\pi^+$ decays using semileptonic $B$ decays}.
\newblock {\em JHEP}, 04:043, 2015.

\bibitem{Aaij:2013ria}
R.~Aaij et~al.
\newblock {Measurements of indirect $CP$ asymmetries in $D^0\to K^-K^+$ and
  $D^0\to\pi^-\pi^+$ decays}.
\newblock {\em Phys. Rev. Lett.}, 112(4):041801, 2014.

\bibitem{Aaij:2013wda}
R.~Aaij et~al.
\newblock {Measurement of $D^0-\bar D^0$ Mixing Parameters and Search for $CP$
  Violation Using $D^0 \to K^+ \pi^-$ Decays}.
\newblock {\em Phys. Rev. Lett.}, 111(25):251801, 2013.

\end{thebibliography}
\endgroup  

\end{document}